\documentclass[12pt,english]{article}
\usepackage[T1]{fontenc}
\usepackage[latin9]{inputenc}
\usepackage[letterpaper]{geometry}
\usepackage{amsthm}
\usepackage{amsmath}
\usepackage{graphicx}
\usepackage{setspace}
\usepackage{amssymb}
\usepackage[authoryear]{natbib}
\makeatletter
\newcommand{\lyxaddress}[1]{
\par {\raggedright #1
\vspace{1.4em}
\noindent\par}
}
\theoremstyle{plain}
\newtheorem{thm}{Theorem}
  \theoremstyle{plain}
  \newtheorem{prop}[thm]{Proposition}
  \theoremstyle{plain}
  \newtheorem{lem}[thm]{Lemma}
 \theoremstyle{definition}
  
  \theoremstyle{plain}
  \newtheorem{cor}[thm]{Corollary}

\DeclareMathOperator{\med}{median}
\DeclareMathOperator{\var}{var}
\DeclareMathOperator{\sk}{skew}

\DeclareMathOperator{\unif}{U}

\makeatother

\usepackage{babel}

\begin{document}

\title{Simple  estimators of false discovery rates given as few as one or
two p-values without strong parametric assumptions\\
\textbf{~}}

\maketitle
~~\\
David R. Bickel

\lyxaddress{Ottawa Institute of Systems Biology\\
Department of Biochemistry, Microbiology, and Immunology\\
University of Ottawa; 451 Smyth Road; Ottawa, Ontario, K1H 8M5}
\begin{abstract}
Multiple comparison procedures that control a family-wise error
rate or false discovery rate provide an achieved error rate as the
adjusted p-value for each hypothesis tested. However, since such p-values
are not probabilities that the null hypotheses are true, empirical
Bayes methods have been devised to estimate such posterior probabilities,
called local false discovery rates (LFDRs) to emphasize the frequency
interpretation of their priors. The main approaches to LFDR estimation,
relying either on numerical algorithms to maximize likelihood or on
the selection of smoothing parameters for nonparametric density estimation,
lack the automatic nature of the methods of error rate control. To
begin filling the gap, this paper introduces automatic methods of
LFDR estimation with proven asymptotic conservatism under the independence
of p-values but without strong parametric assumptions. Simulations
indicate that they remain conservative even for very small numbers
of hypotheses. One of the proposed procedures enables interpreting
the original FDR control rule in terms of LFDR estimation, thereby
facilitating practical interpretation. The most conservative of the
new procedures is applied to measured abundance levels of 20 proteins.

\end{abstract}

\section{\label{sec:Introduction}Introduction}

Since the successful application of the false discovery rate to high-dimensional
biological data \citep{RefWorks:53}, methodological research has
taken two main directions in addition to the hierarchical Bayesian
direction in which a joint prior distribution of all unknown quantities
is given. The purely frequentist line of research has continued to
generalize the theorem of \citet{RefWorks:288} for strictly controlling
the false discovery rate and has resulted in methods of similarly
controlling related quantities such as the number or proportion of
false discoveries. (\citet{dudoit2008multiple} supply a comprehensive
overview of multiple testing in that tradition.) The empirical Bayes
research stream has developed various methods of applying models that
have random parameters as well as unknown, fixed parameters. The hallmark
of the pure frequentist approach to multiple testing, as with frequentism
more generally \citep{Efron1986b}, is the provision of automatic
procedures for data analysis with guarantees regarding their operating
characteristics. In addition, frequentist approaches typically apply
to small numbers of hypotheses as well as to large numbers. By contrast,
the main advantage of the empirical Bayes approach is its ability
to estimate the local counterpart of the false discovery rate, which
is a posterior probability that the null hypothesis is false without
invoking subjective priors. As a posterior probability, the local
false discovery rate is easily interpretable and leads to asymptotically
optimal estimation and prediction; see \citet{efron_large-scale_2010}
for examples. However, that advantage comes at the expense of guaranteed
error rate control and, in the case of nonparametric estimators requiring
the tuning of smoothing parameters, at the expense of automation and
applicability to smaller numbers of hypotheses \citep[e.g., ][]{RefWorks:55}.
Fully parametric methods of estimating the local false discovery rate
tend to require numeric optimization to maximize the likelihood function
\citep[e.g., ][]{ParametricMixtureLFDR,smallScale}. This paper draws
from the strengths of each research direction by proposing an automatic
estimator of the empirical Bayes posterior probability that may be
applied to as few as two hypotheses without making strong parametric
assumptions. 

Some notation will clarify the concepts. In testing $N$ null hypotheses
versus $N$ alternative hypotheses, each of which is either true $\left(A_{i}=1\right)$
or false $\left(A_{i}=0\right)$, the $i$th null hypothesis is considered
\emph{rejected} if the statistic $T_{i}$ falls within some \emph{rejection
region} $\mathcal{T}$. Every rejection is a discovery, a \emph{false
discovery} if the null hypothesis is true $\left(A_{i}=0\right)$
or a \emph{true discovery} otherwise $\left(A_{i}=1\right)$. Thus,
$N_{0}\left(\mathcal{T}\right)$ or $N_{1}\left(\mathcal{T}\right)$,
the number of true or false null hypotheses rejected, is the number
of false or true discoveries, respectively. Then $N_{+}\left(\mathcal{T}\right)=N_{0}\left(\mathcal{T}\right)+N_{1}\left(\mathcal{T}\right)$
is the total number of discoveries \citep{efron_large-scale_2010}.

With the value of each $A_{i}$ unknown but fixed, \citet{RefWorks:288}
defined the \emph{false discovery rate} (FDR) as\[
E\left(\frac{N_{0}\left(\mathcal{T}\right)}{N_{+}\left(\mathcal{T}\right)\vee1}\right),\]
where the denominator is the maximum of $N_{+}\left(\mathcal{T}\right)$
and 1. In other words, the false discovery rate is the expectation
value of the proportion of discoveries that are false with the convention
that the proportion of false discoveries is 0 if no discoveries are
made. While guaranteeing that the FDR does not exceed some critical
level needs that seemingly harmless convention \citep{RefWorks:288},
the convention can cause fatal interpretation problems unless the
probability of making at least one discovery is sufficiently high
\citep{RefWorks:282}. 

A particularly simple and informative alternative to the FDR is the
probability that a null hypothesis is true conditional on its rejection:\[
\Phi\left(\mathcal{T}\right)=\Pr\left(A_{i}=0\vert T_{i}\in\mathcal{T}\right)=\frac{E\left(N_{0}\left(\mathcal{T}\right)\right)}{E\left(N_{+}\left(\mathcal{T}\right)\right)}.\]
Due to its association with Bayes's theorem and its modeling each
$A_{i}$ as a random variable, $\Phi\left(\mathcal{T}\right)$ has
been named the {}``Bayesian false discovery rate'' \citep{RefWorks:54},
a term avoided here since it has conflicting meanings \citep{ISI:000243126900001,ISI:000255959500018}
and since it suggests the fully Bayesian practice of assigning a prior
to every unknown quantity. $\Phi\left(\mathcal{T}\right)$ will be
called the \emph{nonlocal false discovery rate} (NFDR) to distinguish
it from both the FDR and from the \emph{local false discovery rate}
(LFDR), \begin{equation}
\phi\left(t_{i}\right)=\Phi\left(\left\{ t_{i}\right\} \right)=\Pr\left(A_{i}=0\vert T_{i}=t_{i}\right),\label{eq:LFDR}\end{equation}
with $t_{i}$ denoting the observed realization of $T_{i}$. The LFDR
is closer to Bayes-optimal than the NFDR in that it is conditional
on the observed statistic rather than merely on the event that the
statistic lies within $\mathcal{T}$. In addition, the LFDR is intuitively
appealing as the probability that the null hypothesis is true given
the reduced data. 

Thus, the primary reason for introducing conservative methods of NFDR
estimation for as few as a single p-value in Section \ref{sec:Estimation-of-NFDR}
is to repurpose them for conservative LFDR estimation for as few as
two p-values in Section \ref{sec:Estimation-of-LFDR}. Section \ref{sec:Case-study}
features an application to testing 20 hypotheses on the basis of proteomics
data. The simulation study of Section \ref{sec:Simulation-study}
quantifies the performance of three of the new LFDR estimators for
various finite numbers of hypotheses. Finally, Section \ref{sec:Discussion}
provides a brief discussion, and Appendix A collects proofs omitted
from previous sections.

\section{\label{sec:Estimation-of-NFDR}Estimation of nonlocal false discovery
rates}

Let $\pi_{0}=\Pr\left(A_{i}=0\right)$, $\Pi\left(\mathcal{T}\right)=\Pr\left(T_{i}\in\mathcal{T}\right)$,
$\Pi_{0}\left(\mathcal{T}\right)=\Pr\left(T_{i}\in\mathcal{T}\vert A_{i}=0\right)$,
and $\Pi_{1}\left(\mathcal{T}\right)=\Pr\left(T_{i}\in\mathcal{T}\vert A_{i}=1\right)$.
By Bayes's theorem,\begin{equation}
\Phi\left(\mathcal{T}\right)=\Pr\left(A_{i}=0\vert T_{i}\in\mathcal{T}\right)=\frac{\pi_{0}\Pi_{0}\left(\mathcal{T}\right)}{\Pi\left(\mathcal{T}\right)},\label{eq:NFDR}\end{equation}
which is often estimated by substituting 1 for $\pi_{0}$ and $\widehat{\Pi}\left(\mathcal{T};N_{+}\left(\mathcal{T}\right)\right)=N_{+}\left(\mathcal{T}\right)/N$
for $\Pi\left(\mathcal{T}\right)$:\begin{equation}
\widehat{\Phi}\left(\mathcal{T};N_{+}\left(\mathcal{T}\right)\right)=\frac{\Pi_{0}\left(\mathcal{T}\right)}{\widehat{\Pi}\left(\mathcal{T};N_{+}\left(\mathcal{T}\right)\right)}\wedge1,\label{eq:MLE}\end{equation}
the minimum of $\Pi_{0}\left(\mathcal{T}\right)/\widehat{\Pi}\left(\mathcal{T};X\right)$
and 1. If the test statistics are independent of each other, $X=N_{+}\left(\mathcal{T}\right)$
follows the binomial distribution with parameters $N$ and $\Pi\left(\mathcal{T}\right)$,
and $\widehat{\Pi}\left(\mathcal{T}\right)$ is the maximum-likelihood
estimate (MLE) of $\Pi\left(\mathcal{T}\right)$. Thus, $\widehat{\Phi}\left(\mathcal{T}\right)$
is the MLE of $\Pi_{0}\left(\mathcal{T}\right)/\Pi\left(\mathcal{T}\right)$,
which is no less than $\Phi\left(\mathcal{T}\right)$. 

This estimator also provides a convenient statement of the \citet{RefWorks:288}
method of controlling the FDR at level $q$: in terms of upper-tailed
testing, \begin{equation}
\widehat{A}_{i}=\begin{cases}
1 & \text{if }t_{i}\ge t\left(q\right);\\
0 & \text{if }t_{i}<t\left(q\right),\end{cases}\label{eq:control}\end{equation}
where $t\left(q\right)=\inf\left\{ t_{i}:i\in\left\{ 1,\dots,N\right\} ,\widehat{\Phi}\left(\left[t_{i},\infty\right)\right)\le q\right\} $,
$\widehat{A}_{i}=1$ indicates rejection of the $i$th null hypothesis,
and $\widehat{A}_{i}=0$ indicates its acceptance \citep[Corollary 4.2]{efron_large-scale_2010}.
The practical importance of that relationship is discussed in Section
\ref{sec:Discussion}.

The independence model facilitates the derivation of confidence intervals
\citep{efron_large-scale_2010}. For $C\in\left[0,1\right]$ and a
realization $x$ of $X$, let $S_{C}$ and $S_{C}^{-1}$ denote significance
and inverse-significance functions such that \begin{equation}
S_{C}\left(\Pi\left(\mathcal{T}\right);x\right)=\Pr\left(X>x;\Pi\left(\mathcal{T}\right)\right)+C\Pr\left(X=x;\Pi\left(\mathcal{T}\right)\right);\label{eq:binomial-S}\end{equation}
\begin{equation}
S_{C}^{-1}\left(S_{C}\left(\Pi\left(\mathcal{T}\right);x\right);x\right)=\Pi\left(\mathcal{T}\right),\label{eq:inverse-binomial-S}\end{equation}
where $\Pr\left(\bullet;\Pi\left(\mathcal{T}\right)\right)$ denotes
the binomial distribution with parameters $N$ and $\Pi\left(\mathcal{T}\right)$.
Then the standard binomial, one-sided $\left(1-\alpha\right)100\%$
confidence intervals  for $\Pi\left(\mathcal{T}\right)$ \citep{RefWorks:1036}
are $\left[0,S_{0}^{-1}\left(1-\alpha;x\right)\right]$ and $\left[S_{1}^{-1}\left(\alpha;x\right),1\right].$
They are valid confidence intervals:\[
\Pr\left(S_{0}^{-1}\left(1-\alpha;X\right)\ge\Pi\left(\mathcal{T}\right)\right)\ge1-\alpha\]
\begin{equation}
\Pr\left(S_{1}^{-1}\left(\alpha;X\right)\le\Pi\left(\mathcal{T}\right)\right)\ge1-\alpha.\label{eq:confidence}\end{equation}

\begin{figure}
\includegraphics[scale=0.5]{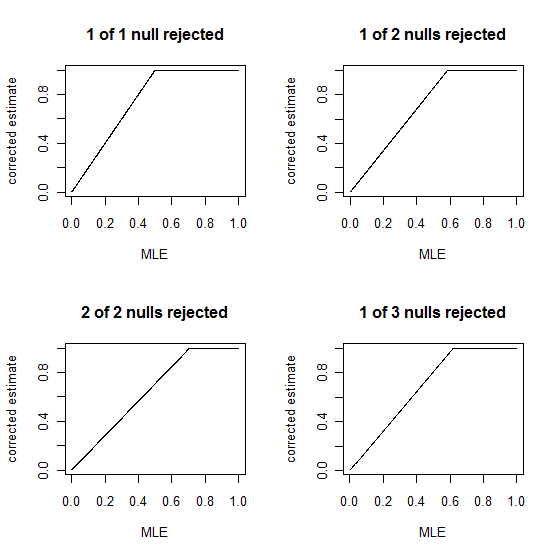}

\caption{Nonlocal false discovery rate estimates. The \textbf{corrected estimate}
is $\widetilde{\Phi}\left(\mathcal{T};x\right)$, the confidence posterior
median, and the \textbf{MLE} is $\widehat{\Phi}\left(\mathcal{T};x\right)$,
the maximum likelihood estimate. \label{fig:Nonlocal-FDR-estimates}}

\end{figure}

\begin{figure}
\includegraphics{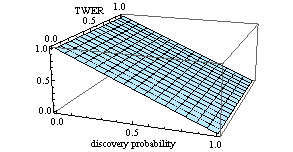}\includegraphics{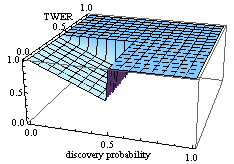}

\caption{Probability that the \textbf{MLE} $\widehat{\Phi}\left(\mathcal{T};x\right)$
{[}left{]} and the \textbf{corrected estimate} $\widetilde{\Phi}\left(\mathcal{T};x\right)$
{[}right{]} is at least as high as the upper bound of the nonlocal
false discovery rate when $N=1$. Here, {}``TWER'' is the Type I
test-wise error rate, and {}``discovery probability'' is the probability
of rejecting any given null hypothesis.\label{fig:probGE1}}

\end{figure}

\begin{figure}
\includegraphics{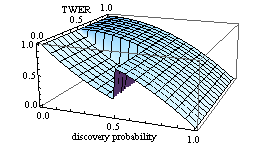}\includegraphics{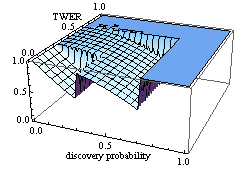}

\caption{Probability that the \textbf{MLE} $\widehat{\Phi}\left(\mathcal{T};x\right)$
{[}left{]} and the \textbf{corrected estimate} $\widetilde{\Phi}\left(\mathcal{T};x\right)$
{[}right{]} is at least as high as the upper bound of the nonlocal
false discovery rate when $N=2$. Here, {}``TWER'' is the Type I
test-wise error rate, and {}``discovery probability'' is the probability
of rejecting any given null hypothesis.\label{fig:probGE2}}

\end{figure}

Since $A_{i}$ rather than the uncertainty of $\Phi\left(\mathcal{T}\right)$
is of direct interest, the main value of the confidence intervals
is in the construction of better point estimates of $\Phi\left(\mathcal{T}\right)$
and thus of $1-A_{i}$ for all $i$ satisfying $T_{i}\in\mathcal{T}$.
A point estimate $\Phi^{\ast}\left(\mathcal{T};x\right)$ that satisfies
$\Pr\left(\Phi^{\ast}\left(\mathcal{T};x\right)\ge\Phi\left(\mathcal{T}\right)\right)\ge1/2$
for all $\pi_{0},\Pi_{0}\left(\mathcal{T}\right),\Pi\left(\mathcal{T}\right)\in\left[0,1\right]$
is called a \emph{median conservative estimator} of $\Phi\left(\mathcal{T}\right)$.
According to the following proposition, one such estimator is $\widetilde{\Phi}\left(\mathcal{T};x\right)=\widetilde{\Phi}_{1}\left(\mathcal{T};x\right),$
the $C=1$ special case of\begin{equation}
\widetilde{\Phi}_{C}\left(\mathcal{T};x\right)=\frac{\Pi_{0}\left(\mathcal{T}\right)}{S_{C}^{-1}\left(1/2;x\right)}\wedge1.\label{eq:posterior-median}\end{equation}
Each $\widetilde{\Phi}_{C}\left(\mathcal{T};x\right)$ is called a
\emph{confidence-posterior median }of $\Phi\left(\mathcal{T}\right)$
since it is a median of $\Phi\left(\mathcal{T}\right)$ considered
as a function of a random binomial parameter of distribution function
$S_{C}^{-1}\left(\bullet;x\right)$ \citep{CoherentFrequentism}.
$\widetilde{\Phi}\left(\mathcal{T};x\right)$ may be considered as
a conservative correction to the MLE, as seen in Fig. \ref{fig:Nonlocal-FDR-estimates}.
\begin{prop}
\label{pro:conservative-NFDR}Under the independence of $T_{1},...,T_{N}$,
the random quantity $\widetilde{\Phi}\left(\mathcal{T};X\right)$
is a median conservative\emph{ }estimator of $\Phi\left(\mathcal{T}\right)$.\end{prop}
\begin{proof}
Independence entails equation \eqref{eq:confidence}, which implies
that\[
\Pr\left(\frac{\Pi_{0}\left(\mathcal{T}\right)}{S_{1}^{-1}\left(1/2;X\right)}\ge\frac{\Pi_{0}\left(\mathcal{T}\right)}{\Pi\left(\mathcal{T}\right)}\right)=\Pr\left(S_{1}^{-1}\left(1/2;X\right)\le\Pi\left(\mathcal{T}\right)\right)\ge1/2.\]
Since, by formula \eqref{eq:NFDR}, $\Phi\left(\mathcal{T}\right)\le\Pi_{0}\left(\mathcal{T}\right)/\Pi\left(\mathcal{T}\right)$
and since $\Pi_{0}\left(\mathcal{T}\right)/\Pi\left(\mathcal{T}\right)\le1$,
it follows from equation \eqref{eq:posterior-median} that $\Phi_{1/2}\left(\mathcal{T};X\right)$
is median conservative:\begin{equation}
\Pr\left(\widetilde{\Phi}\left(\mathcal{T};X\right)\ge\Phi\left(\mathcal{T}\right)\right)\ge1/2.\label{eq:median-conservative}\end{equation}

\end{proof}
 That the MLE is not median conservative is evident from the left-hand
sides of Figs. \ref{fig:probGE1}-\ref{fig:probGE2} for the $N=1,2$
cases: $\Pr\left(\widehat{\Phi}\left(\mathcal{T}_{\alpha};X\right)\ge\Phi\left(\mathcal{T}\right)\right)<1/2$
for some combinations of the test-wise error rate $\alpha$ and the
discovery probability $\Pi\left(\mathcal{T}_{\alpha}\right)$, where
$\mathcal{T}_{\alpha}$ is a level-$\alpha$ critical region such
that $\Pi_{0}\left(\mathcal{T}_{\alpha}\right)=\alpha$. For contrast
with the corrected estimates given by the confidence-posterior median
$\widetilde{\Phi}\left(\mathcal{T};x\right)$, the right-hand sides
of Figs. \ref{fig:probGE1}-\ref{fig:probGE2} illustrate formula
\eqref{eq:median-conservative}, also for $N=1,2$. Accordingly, $\widetilde{\Phi}\left(\mathcal{T};X\right)$
will be called the \emph{corrected estimate of the NFDR}. 

The expectation value of a random quantity with respect to $S_{C}\left(\bullet;x\right)$
as the distribution function of the random binomial parameter is called
a \emph{confidence-posterior mean}. For example, writing $\Pi^{\prime}$
as the dummy variable of integration, the confidence-posterior mean
of $\Pi\left(\mathcal{T}\right)$ is $\int\Pi^{\prime}dS_{C}\left(\Pi^{\prime};x\right)$.
Likewise, according to equation \eqref{eq:NFDR}, the confidence-posterior
mean\emph{ }of $\Phi\left(\mathcal{T}\right)$, \begin{equation}
\bar{\Phi}_{C}\left(\mathcal{T};x,\pi_{0}\right)=\int\left(\frac{\pi_{0}\Pi_{0}\left(\mathcal{T}\right)}{\Pi^{\prime}}\right)dS_{C}\left(\Pi^{\prime};x\right),\label{eq:confidence-posterior-mean}\end{equation}
is a Bayes-confidence-posterior probability that $A_{i}=0$ given
$T_{i}\in\mathcal{T}$. As such, it rivals the hierarchical Bayes
approach to accounting for the uncertainty in $\Phi\left(\mathcal{T}\right)$
and is complete with a decision theory based on minimizing expected
loss \citep{conditional2009,CoherentFrequentism}, and yet without
requiring a hyperprior distribution. In practice, $\pi_{0}$ will
again be set to 1, yielding $\bar{\Phi}_{C}\left(\mathcal{T};x\right)=\bar{\Phi}_{C}\left(\mathcal{T};x,1\right)$,
the confidence-posterior mean\emph{ }of $\Pi_{0}\left(\mathcal{T}\right)/\Pi\left(\mathcal{T}\right)$,
as an upper bound of the confidence-posterior mean\emph{ }of $\Phi\left(\mathcal{T}\right)$. 

While median conservatism is a finite-$N$ property, concepts of asymptotic
conservatism become prominent in the results of the next section.
A random variable $\widehat{\gamma}\left(X\right)$ is a \emph{conservative
estimator} of some constant $\gamma$ if $\lim_{N\rightarrow\infty}\Pr\left(\widehat{\gamma}\left(X\right)\ge\gamma\right)=1$.
Likewise, $\widehat{\gamma}\left(X\right)$ is a \emph{conservative
predictor} of some random variable $\gamma\left(X\right)$ if $\lim_{N\rightarrow\infty}\Pr\left(\widehat{\gamma}\left(X\right)\ge\gamma\left(X\right)\right)=1$.
The estimators of the NFDR considered above are conservative, as will
be proven in Appendix A:
\begin{lem}
\label{lem:asymptotic}If $T_{1},...,T_{N}$ are IID, then $\widehat{\Phi}\left(\mathcal{T};X\right)$,
the members of $\left\{ \widetilde{\Phi}_{C}\left(\mathcal{T};X\right):C\in\left[0,1\right]\right\} $
, and the members of $\left\{ E_{C}\left(\Phi\left(\mathcal{T}\right);X,1\right):C\in\left[0,1\right]\right\} $
are conservative estimators of $\Phi\left(\mathcal{T}\right)$.
\end{lem}

\section{\label{sec:Estimation-of-LFDR}Estimation of local false discovery
rates}

\subsection{Additional notation}

Let $\mathfrak{p}$ designate a one-to-one, monotonic map from each
statistic to a p-value such that $p_{i}=\mathfrak{p}\left(t_{i}\right)$
is the p-value that corresponds to the $i$th null hypothesis, which
would be rejected if $p_{i}\le\alpha$ for some Type I test-wise error
rate $\alpha\in\left[0,1\right]$. Thus, $\mathcal{T}_{\alpha}$ is
constrained to satisfy $\mathcal{T}_{\alpha}=\left\{ t:\mathfrak{p}\left(t\right)\le\alpha\right\} $.
(The requirement that $\mathfrak{p}$ be invertible does not rule
out two-sided tests since they can be equivalently formulated as one-sided
tests by transforming the test statistic. A two-sided \emph{t}-test
will be used in Section \ref{sec:Case-study}.) 

Denote the random p-value of the $i$th null hypothesis by $P_{i}=\mathfrak{p}\left(T_{i}\right)$.
The order statistics of $p_{1},\dots,p_{N}$ and $P_{1},\dots,P_{N}$
are $p_{\left(1\right)},\dots,p_{\left(N\right)}$ and $P_{\left(1\right)},\dots,P_{\left(N\right)}$,
respectively. In the same way, $r_{i}$ is the rank of $p_{i}$ among
the other observed p-values, and $R_{i}$ is the rank of $P_{i}$
among the other random p-values. The presentation of the methodology
is simplified by ensuring that ties do not occur in $p_{1},\dots,p_{N}$,
achievable by breaking ties with a pseudorandom-number generator,
and that they occur with probability 0 in $P_{1},\dots,P_{N}$, which
follows from the stipulations that $T_{i}$ be a continuous random
variable and that the $T_{1},\dots,T_{N}$ be IID. Hence, $p_{\left(r_{i}\right)}=p_{i}$
and $\Pr\left(P_{\left(R_{i}\right)}=P_{i}\right)=1$ for all $i=1,\dots,N$.

For economy of notation, $\Phi\left(\alpha\right)=\Phi\left(\mathcal{T}_{\alpha}\right)$
and \[
\varphi\left(p\right)=\Pr\left(A_{i}=0\vert\mathfrak{p}\left(T_{i}\right)=p\right)\]
respectively denote the NFDR and, for any $p\in\left[0,1\right]$,
the LFDR. Since $t_{i}=\mathfrak{p}^{-1}\left(p_{i}\right)$ for any
$i=1,\dots,N$, each LFDR agrees with equation \eqref{eq:LFDR}: $\varphi\left(p_{i}\right)=\phi\left(t_{i}\right)$.
Similarly, $N_{j}\left(\alpha\right)=N_{j}\left(\mathcal{T}_{\alpha}\right)\,\left[j=0,1\right]$
and $N_{+}\left(\alpha\right)=N_{+}\left(\mathcal{T}_{\alpha}\right)$
are the conditional and marginal numbers of discoveries, and $\Pi_{j}\left(\alpha\right)=\Pi_{j}\left(\mathcal{T}_{\alpha}\right)\,\left[j=0,1\right]$
and $\Pi_{+}\left(\alpha\right)=\Pi_{+}\left(\mathcal{T}_{\alpha}\right)$
are the conditional and marginal probabilities that $T_{i}\in\mathcal{T}_{\alpha}$.
Lastly, $\Phi^{\ast}\left(\alpha\right)=\Phi^{\ast}\left(\alpha;N_{+}\left(\alpha\right)\right)=\Phi^{\ast}\left(\mathcal{T}_{\alpha};N_{+}\left(\mathcal{T}_{\alpha}\right)\right)$
will represent an estimate of the NFDR, where the function $\Phi^{\ast}$
may be $\widehat{\Phi}$, $\widetilde{\Phi}_{C}$, or $\bar{\Phi}_{C}$.

\subsection{Conservative LFDR estimation}

The LFDR $\varphi\left(p_{i}\right)$ will be estimated by the NFDR
estimated with $\alpha$ equal to the p-value of twice the rank of
$p_{i}$ if possible or estimated by 1 otherwise. That is, given $\Phi^{\ast}$
as the estimator of the NFDR, $\varphi\left(p_{i}\right)$ is estimated
by \[
\varphi\left(r_{i};\Phi^{\ast}\right)=\begin{cases}
\Phi^{\ast}\left(p_{\left(2r_{i}\right)};N_{+}\left(p_{\left(2r_{i}\right)}\right)\right) & \text{if }r_{i}\le\frac{N}{2};\\
1 & \text{if }r_{i}>\frac{N}{2}.\end{cases}\]
For example, the MLE, the corrected estimate, and the confidence-mean
estimates of the LFDR are \begin{equation}
\widehat{\varphi}\left(r_{i}\right)=\varphi\left(r_{i};\widehat{\Phi}\right)=\widehat{\Phi}\left(p_{\left(2r_{i}\right)}\right);\label{eq:LFDR-MLE}\end{equation}
\begin{equation}
\widetilde{\varphi}\left(r_{i}\right)=\varphi\left(r_{i};\widetilde{\Phi}\right)=\widetilde{\Phi}\left(p_{\left(2r_{i}\right)}\right);\label{eq:corrected-estimate}\end{equation}
\[
\bar{\varphi}_{C}\left(r_{i}\right)=\varphi\left(r_{i};\bar{\Phi}_{C}\right)=\bar{\Phi}_{C}\left(p_{\left(2r_{i}\right)}\right)\]
for any $r_{i}\le N/2$ and $\widehat{\varphi}\left(r_{i}\right)=\widetilde{\varphi}\left(r_{i}\right)=\bar{\varphi}_{C}\left(r_{i}\right)=1$
for any $r_{i}>N/2$.

The theorem stated below establishes a sense in which an LFDR estimator
is conservative under general assumptions, including one involving
the following conditional version of a definition of skewness attributed
to Karl Pearson \citep{ISI:000227879800009}. The \emph{Pearson skewness}
of a random variable $Y$, conditional on event $\mathcal{E}$ is
\[
\sk\left(Y\vert\mathcal{E}\right)=3\frac{E\left(Y\vert\mathcal{E}\right)-\med\left(Y\vert\mathcal{E}\right)}{\sqrt{\var\left(Y\vert\mathcal{E}\right)}}.\]
Let $f$ and $F$ respectively denote the probability density and
cumulative distribution functions of $P_{i}$ for each $i=1,\dots,N$.
\begin{thm}
\label{thm:conservatism}Assume that $T_{1},...,T_{N}$ are continuous
and IID and that $\varphi:\left[0,1\right]\rightarrow\left[0,1\right]$
is monotonically nondecreasing. If $\Phi^{\ast}\left(\alpha\right)$
is a conservative estimator of $\Phi\left(\alpha\right)$ and $\sk\left(\varphi\left(P_{i}\right)\vert P_{i}\le\alpha\right)\ge0$
for any $\alpha\in\left(0,1\right]$ , then \textup{$\varphi\left(R_{i};\Phi^{\ast}\right)$
is a conservative predictor of $\varphi\left(P_{i}\right)$.}

\end{thm}
The proof will appear in Appendix A. \citet{ISI:A1997XM80000002}
reviewed various sets of sufficient conditions for $E\left(Y\right)\ge\med\left(Y\right)$
(nonnegative Pearson skewness, $\sk\left(Y\right)\ge0$). This corollary
of the theorem follows readily from Lemma \ref{lem:asymptotic}:
\begin{cor}
Under the conditions of Theorem \ref{thm:conservatism}, $\widehat{\varphi}\left(R_{i}\right)$,
the members of $\left\{ \widetilde{\varphi}_{C}\left(R_{i}\right):C\in\left[0,1\right]\right\} $,
and the members of $\left\{ \bar{\varphi}_{C}\left(R_{i}\right):C\in\left[0,1\right]\right\} $
are conservative predictors of $\varphi\left(P_{i}\right)$.
\end{cor}
Stated less formally, the proposed maximum-likelihood LFDR estimate,
corrected LFDR estimate, and bound on the confidence-posterior-mean
LFDR conservatively estimate the LFDR given a sufficiently large number
of hypotheses.

Although $\varphi=\varphi\left(\bullet\right)$ is monotonically increasing,
$\varphi\left(\bullet;\Phi^{\ast}\right)$ in general is not: LFDR
estimates do not necessarily preserve the order of the p-values, which
is the order of the actual LFDRs. Thus, in the next two sections,
the monotonicity of the estimates, \begin{equation}
\varphi\left(r_{1};\Phi^{\ast}\right)\le\dots\le\varphi\left(r_{N};\Phi^{\ast}\right),\label{eq:monotonicity}\end{equation}
is enforced by this algorithm used with step-down multiple comparison
procedures \citep{RefWorks:275,dudoit2008multiple}: $\varphi\left(r_{2};\Phi^{\ast}\right)$
is changed to $\varphi\left(r_{1};\Phi^{\ast}\right)$ if $\varphi\left(r_{2};\Phi^{\ast}\right)<\varphi\left(r_{1};\Phi^{\ast}\right)$,
then $\varphi\left(r_{3};\Phi^{\ast}\right)$ is changed to $\varphi\left(r_{2};\Phi^{\ast}\right)$
if $\varphi\left(r_{3};\Phi^{\ast}\right)<\varphi\left(r_{2};\Phi^{\ast}\right)$,
etc. Since such monotonicity enforcement cannot decrease the estimates,
conservatism is maintained.

\section{\label{sec:Case-study}Application to proteomics data}

Levels of 20 proteins were measured in 90 women with breast cancer
(55 HER2-positive and 35 mostly ER/PR-positive) and a group of 64
healthy women. (The data \citep{ProData2009b} are from Alex Miron\textquoteright{}s
lab at the Dana-Farber Cancer Institute.) To approximate normality,
the abundance levels were transformed first by adding the 25th percentile
over all the proteins and over all the healthy women (yielding positive
levels without a hard threshold) and then by taking the logarithm. 

\begin{figure}
\includegraphics[scale=0.5]{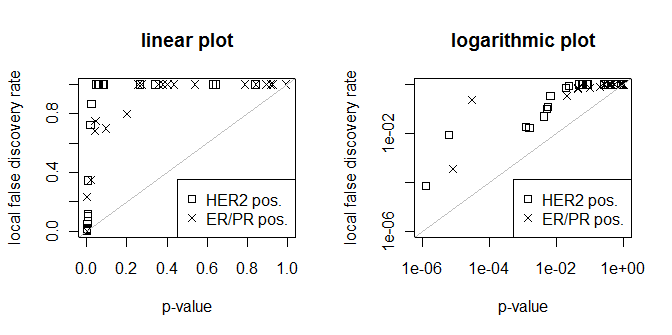}

\caption{Corrected estimate of the local false discovery rate versus the p-value
for the application to proteomics data.\label{fig:Cancer}}

\end{figure}

For each cancer group (HER2 and ER/PR), there are 20 null hypotheses
of no mean difference in the transformed level between cancer and
healthy groups. Fig. \ref{fig:Cancer} plots $\widetilde{\varphi}\left(r_{i}\right)$
against $p_{i}$, the p-value of the two-sample t-test with equal
variances for the null hypothesis that the $i$th protein has the
same expected abundance level in a cancer group as in the healthy
group. Each displayed estimate of the LFDR is easily interpretable
as a conservative estimate of the posterior probability that a given
protein has the same average level of abundance in a cancer group
as it does in the control group.

\section{\label{sec:Simulation-study}Simulation study}

Let $\chi_{1,\delta}^{2}$ denote the noncentral $\chi^{2}$ distribution
with 1 degree of freedom and with noncentrality parameter $\delta$.
Many test statistics are asymptotically $\chi_{1,\delta}^{2}$, with
$\delta=0$ under the null hypothesis and $\delta>0$ under any alternative
hypothesis. Notable tests with such statistics are the likelihood
ratio test with a scalar parameter of interest and a local alternative
hypothesis and two-sided tests with asymptotically normal statistics
of unit variance. As a result, this limit is highly relevant to problems
in modern biology~\citep{smallScale}, including that of Section \ref{sec:Case-study}.

\begin{figure}
\includegraphics[scale=0.5]{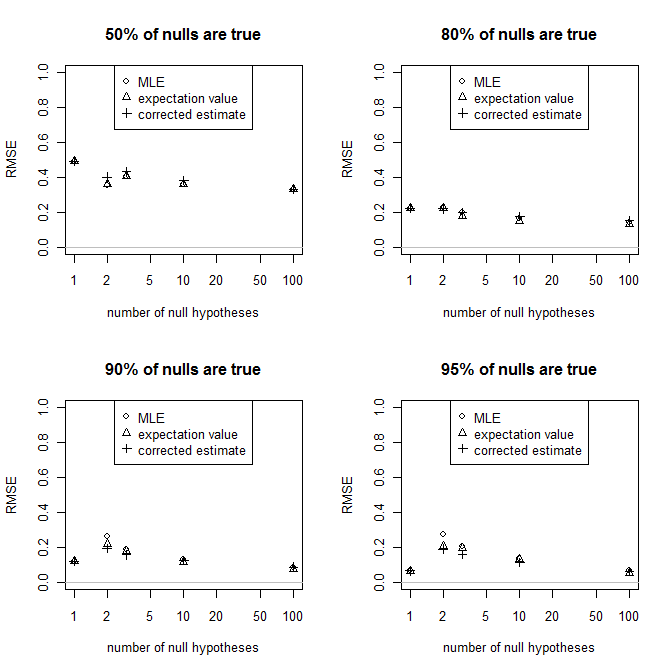}

\caption{Root mean squared error in local false discovery rate estimation versus
$N$. Each of the four panels corresponds to a different value of
$\pi_{0}$.\label{fig:Root-mean-squared-error-2}}

\end{figure}

\begin{figure}
\includegraphics[scale=0.5]{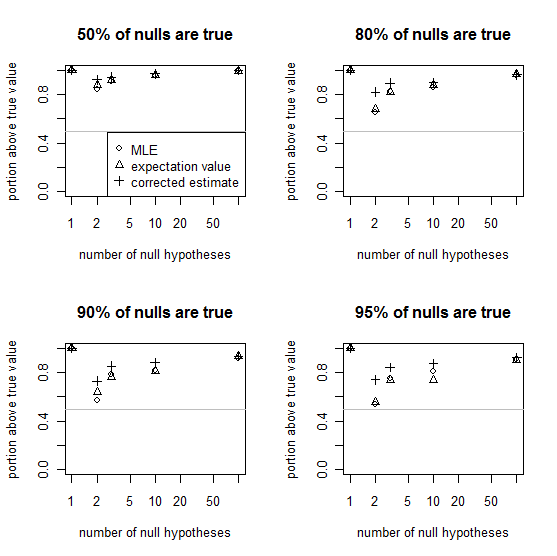}

\caption{Conservatism in local false discovery rate estimation versus $N$.
Conservatism is measured by the proportion of estimates that exceed
the local false discovery rates they estimate. Each of the four panels
corresponds to a different value of $\pi_{0}$.\label{fig:Proportion}}

\end{figure}

\begin{figure}
\includegraphics[scale=0.5]{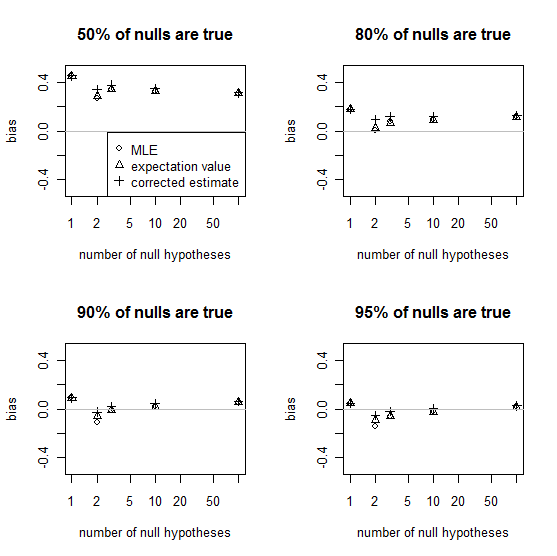}\caption{Arithmetic bias in local false discovery rate estimation versus $N$.
Each of the four panels corresponds to a different value of $\pi_{0}$.\label{fig:Bias}}

\end{figure}

Consequently, each simulated data set consisted of $N$ test statistics
independently drawn from $\chi_{1,0}^{2}$ with probability $\pi_{0}$
and from $\chi_{1,2}^{2}$ with probability $1-\pi_{0}$ for each
of four values of $\pi_{0}$ and for each of five values of $N$.
For every $\left\langle \pi_{0},N\right\rangle $ configuration, 100
independent data sets were generated. In Figs. \ref{fig:Root-mean-squared-error-2},
\ref{fig:Proportion}, and \ref{fig:Bias}, the LFDR estimates are
based on NFDR estimates for $i=1,\dots,N$: the {}``MLE'' $\widehat{\varphi}\left(r_{i}\right)$
estimates the NFDR by equation \eqref{eq:LFDR-MLE}, the {}``expectation
value'' $\bar{\varphi}_{1/2}\left(r_{i}\right)$ is based on the
$\pi_{0}=1$ upper bound of the confidence-posterior mean of the NFDR
\eqref{eq:confidence-posterior-mean} with $C=1/2$, and the {}``corrected
estimate'' $\widetilde{\varphi}\left(r_{i}\right)$ is based on the
upper confidence-posterior median of the NFDR. (Approximation of each
value of $\bar{\varphi}_{1/2}\left(r_{i}\right)$ was achieved by
drawing 100 independent Monte Carlo samples from $S_{1/2}\left(\bullet;x\right)$
via $S_{1/2}^{-1}\left(U;x\right),U\sim\unif\left(0,1\right)$ and
by averaging according to equation \eqref{eq:confidence-posterior-mean}.)\\

\section{\label{sec:Discussion}Discussion}

Compared to previous estimators of the LFDR, the main advantages of
the proposed methods are their proven conservatism (Theorem \ref{thm:conservatism})
and their applicability to very small numbers of hypotheses without
strong parametric assumptions. The algorithms are simple, requiring
neither numeric likelihood maximization nor nonparametric smoothing
procedures. The algorithm for $\widehat{\varphi}\left(r_{i}\right)$,
the proposed MLE, is particularly simple, being only slightly more
complicated than the FDR-controlling procedure of \citet{RefWorks:288}.

In fact, $\widehat{\varphi}\left(r_{i}\right)$ can shed light on
the practical interpretation of applications of that FDR procedure.
From equations \eqref{eq:control} and \eqref{eq:LFDR-MLE}, it can
be seen that the value $q$ at which the FDR is controlled for a set
of rejected null hypotheses is equal to $\widehat{\varphi}\left(r_{i\left(q\right)}\right)$
when violations of monotonicity \eqref{eq:monotonicity} are neglected,
where $i\left(q\right)$ is the index such that $p_{i\left(q\right)}$
is the p-value equal to the median of the p-values in the rejection
set. Since $\widehat{\varphi}\left(r_{i\left(q\right)}\right)=q$
is simply a conservative estimate of the LFDR corresponding to that
median p-value, the lowest half of the p-values of the hypotheses
rejected by the \citet{RefWorks:288} procedure have conservatively
estimated posterior probabilities of truth less than or equal to $q$. 

While Theorem \ref{thm:conservatism} guarantees conservative performance
only for sufficiently large numbers of hypotheses, examples of finite-$N$
applications were provided in the proteomics case study and in the
simulation study. That the proposed methods conservatively estimate
the LFDR is evident from the proportion of estimates exceeding the
true value (Fig. \ref{fig:Proportion}). The slightly negative arithmetic
bias sometimes seen (Fig. \ref{fig:Bias}) results from forbidding
estimates from exceeding 100\% rather than from any anti-conservatism.
Fig. \ref{fig:Root-mean-squared-error-2} illustrates how the overall
performance of the estimators, owing to their conservative nature,
perform better for higher proportions of true null hypotheses.

\section*{Acknowledgments}

The \texttt{Biobase} \citep{RefWorks:161} package of \texttt{R} \citep{R2008}
facilitated the computations.This research was partially supported  by the Canada Foundation for
Innovation, by the Ministry of Research and Innovation of Ontario,
and by the Faculty of Medicine of the University of Ottawa. 

\begin{flushleft}
\bibliographystyle{elsarticle-harv}
\bibliography{refman}

\par\end{flushleft}

\newpage{}

\section*{\label{sec:Proofs-of-theorems}Appendix A: Additional proofs}

\subsection*{Proof of Lemma \ref{lem:asymptotic}}

By the definition of a conservative estimator and by equation \eqref{eq:NFDR},
any random variable of the form\[
\Phi^{\ast}\left(\mathcal{T};X\right)=\frac{\Pi_{0}\left(\mathcal{T}\right)}{\Pi^{\ast}\left(\mathcal{T};X\right)}\wedge1\]
is a conservative estimator of $\Phi\left(\mathcal{T}\right)$ if
the random variable $\Pi^{\ast}\left(\mathcal{T};X\right)$ converges
to $\Pi\left(\mathcal{T}\right)$ in probability since $\pi_{0}\le1$.
The estimators $\widehat{\Phi}\left(\mathcal{T};X\right)$ and, for
any $C\in\left[0,1\right]$, $\widetilde{\Phi}_{C}\left(\mathcal{T};X\right)$
are of that form with $\Pi^{\ast}\left(\mathcal{T};X\right)=\widehat{\Pi}\left(\mathcal{T};X\right)$
and $\Pi^{\ast}\left(\mathcal{T};X\right)=S_{C}^{-1}\left(1/2;X\right)$,
respectively. The convergence of $\widehat{\Pi}\left(\mathcal{T};X\right)$
to $\Pi\left(\mathcal{T}\right)$ is guaranteed by the weak law of
large numbers. Since $S_{C}^{-1}\left(1/2;X\right)$ is the median
of the random variable that has $S_{C}\left(\bullet;x\right)$ as
its distribution function and since $S_{C}\left(\bullet;x\right)$
is an asymptotic confidence distribution in the sense of \citet{RefWorks:1037},
a sufficient condition for its convergence to $\Pi\left(\mathcal{T}\right)$
is that fixed-level confidence intervals formed by $S_{C}\left(\bullet;x\right)$
degenerate to a point as $N\rightarrow\infty$ \citep[Theorem 3.1]{RefWorks:1037}.
That condition is met since $S_{C}\left(\bullet;x\right)$  is defined
by equation \eqref{eq:binomial-S}, consistent with the confidence
intervals of \citet{RefWorks:1036}. Thus, the conservatism of $\widehat{\Phi}\left(\mathcal{T};X\right)$
and $\widetilde{\Phi}_{C}\left(\mathcal{T};X\right)$ are established.

Similarly, because $\Pi_{0}\left(\mathcal{T}\right)/\Pi\left(\mathcal{T}\right)\ge\Phi\left(\mathcal{T}\right)$,
the conservatism of $\bar{\Phi}_{C}\left(\mathcal{T};X,1\right)$
follows from its convergence to $\Pi_{0}\left(\mathcal{T}\right)/\Pi\left(\mathcal{T}\right)$
in probability. Since $\bar{\Phi}_{C}\left(\mathcal{T};X,1\right)$
as defined in equation \eqref{eq:confidence-posterior-mean} is a
confidence posterior mean of $\Pi_{0}\left(\mathcal{T}\right)/\Pi\left(\mathcal{T}\right)$,
its convergence to $\Pi_{0}\left(\mathcal{T}\right)/\Pi\left(\mathcal{T}\right)$
follows from the two conditions of \citet[Theorem 3.2]{RefWorks:1037}:
\begin{enumerate}
\item fixed-level confidence intervals formed by the asymptotic confidence
distribution of $\Pi_{0}\left(\mathcal{T}\right)/\Pi\left(\mathcal{T}\right)$
degenerate to a point as $N\rightarrow\infty$;
\item the confidence posterior variance\[
\int\left[\left(\frac{\Pi_{0}\left(\mathcal{T}\right)}{\Pi^{\prime}}\right)-\bar{\Phi}_{C}\left(\mathcal{T};X,1\right)\right]^{2}dS_{C}\left(\Pi^{\prime};X\right),\]
is bounded in probability. 
\end{enumerate}
The first condition results from the monotonicity between $\Pi_{0}\left(\mathcal{T}\right)/\Pi^{\prime}$
and $\Pi^{\prime}$ in the integrand of equation \eqref{eq:confidence-posterior-mean},
in which $\Pi_{0}\left(\mathcal{T}\right)$ is fixed, and the fact
that, as argued above to establish the conservatism of $\widetilde{\Phi}_{C}\left(\mathcal{T};X\right)$,
the degeneracy condition is met for $S_{C}\left(\bullet;x\right)$,
the asymptotic confidence distribution of $\Pi\left(\mathcal{T}\right)$.
The second condition follows trivially from the fact that the domain
of $S_{C}$ is $\left[0,1\right]$, thereby establishing the conservatism
of $\bar{\Phi}_{C}\left(\mathcal{T};X,1\right)$.

\subsection*{Proof of Theorem \ref{thm:conservatism}}

Since $\Phi\left(\alpha\right)=E\left(\varphi\left(P_{i}\right)\vert P_{i}\le\alpha\right)$,
the nonnegative-skewness condition implies \[
\Phi\left(\alpha\right)\ge\med\left(\varphi\left(P_{i}\right)\vert P_{i}\le\alpha\right).\]
Thus, defining the variables $P_{i}^{\prime}$ and $P_{\left(i\right)}^{\prime}$
to be IID with $P_{i}$ and $P_{\left(i\right)}$, respectively, for
$i=1,\dots,N$, \[
\Phi\left(P_{\left(2R_{i}\right)}\right)\ge\med\left(\varphi\left(P_{i}^{\prime}\right)\vert P_{i}^{\prime}\le P_{\left(2R_{i}\right)}\right)\]
almost surely. The monotonicity of $\varphi$ implies that, almost
surely,\[
\med\left(\varphi\left(P_{i}^{\prime}\right)\vert P_{i}^{\prime}\le P_{\left(2R_{i}\right)}\right)=\med\left(\varphi\left(P_{i}^{\prime}\right)\vert\varphi\left(P_{i}^{\prime}\right)\le\varphi\left(P_{\left(2R_{i}\right)}\right)\right);\]
\[
\lim_{N\rightarrow\infty}\Pr\left(\med\left(\varphi\left(P_{i}^{\prime}\right)\vert P_{i}^{\prime}\le P_{\left(2R_{i}\right)}\right)=\varphi\left(P_{\left(R_{i}\right)}\right)\right)=1.\]
Because the conservatism of $\Phi^{\ast}\left(\alpha\right)$ means
$\lim_{N\rightarrow\infty}\Pr\left(\Phi^{\ast}\left(\alpha\right)\ge\Phi\left(\alpha\right)\right)=1,$\begin{eqnarray*}
1 & = & \lim_{N\rightarrow\infty}\Pr\left(\Phi^{\ast}\left(P_{\left(2R_{i}\right)}\right)\ge\med\left(\varphi\left(P_{i}^{\prime}\right)\vert P_{i}^{\prime}\le P_{\left(2R_{i}\right)}\right)\right)\\
 & = & \lim_{N\rightarrow\infty}\Pr\left(\Phi^{\ast}\left(P_{\left(2R_{i}\right)}\right)\ge\varphi\left(P_{\left(R_{i}\right)}\right)=\varphi\left(P_{i}\right)\right).\end{eqnarray*}

\subsection*{}
\end{document}